\documentclass[a4paper]{VisionStyle}
\usepackage{epsfig}

\begin{document}

\title{First deep XMM-Newton observations of the LMC: Identifying LMC 
intrinsic source populations}

\author{F.\,Haberl} 

\institute{
  Max-Planck-Institut f\"ur extraterrestrische Physik,
  Giessenbachstra{\ss}e 1, 85748 Garching, Germany
          }

\maketitle 

\begin{abstract}

The first X-ray survey of the Large Magellanic Cloud (LMC) was
performed with the Einstein satellite, revealing diffuse X-ray
emission from hot gas and discrete X-ray sources. The ROSAT
observations between 1990 and 1998 supplied the most sensitive
survey with imaging instruments in the soft X-ray band (0.1 - 2.4 keV).
The PSPC and HRI observations covered 59 square degrees of the LMC and
yielded a catalogue of about 1000 sources.
 
Large efforts were undertaken to identify and classify the X-ray sources
according to the nature of their X-ray emission. X-ray properties
were used together with information from other electro-magnetic
wavelength bands to screen out foreground stars and background
objects from the LMC intrinsic X-ray source population which
comprises X-ray binaries, supernova remnants and supersoft sources.
However, the vast majority of sources still remains of unknown nature.
 
First deep XMM-Newton observations of selected regions in the LMC
demonstrate a large step forward in the identification of LMC X-ray
sources. The large collecting area together with imaging detectors
covering energies 0.1 - 15 keV with far improved spectral resolution
allows to determine the nature of an object directly from the X-ray
properties down to low flux levels of a few 10$^{-14}$ erg cm$^{-2}$ s$^{-1}$.
First results of a deep observation north of LMC X-4 are presented,
which reveal the presence of new supernova remnants and X-ray binaries.                                  

\end{abstract}

\section{Introduction}
  
First X-ray surveys of the Large Magellanic Cloud with imaging instruments were
performed by the Einstein Observatory between 1997 and 1981 (\cite{fhaberl-E3:lo81},
\cite{fhaberl-E3:wa91}) and resulted in the discovery of diffuse X-ray emission
from hot interstellar gas and a catalogue of about 100 discrete X-ray sources.
The most sensitive and most complete survey was carried out by ROSAT from 1990 to 1998
in the energy range 0.1--2.4 keV. The large field of view of the ROSAT PSPC detector
yielded a large coverage of the LMC area by combining the data from hundreds of 
pointings to various targets in the LMC (\cite{fhaberl-E3:hp99a}). 
Due to monitoring programs some fields were
covered more than 30 times during the lifetime of ROSAT. One such region is located
between the high mass X-ray binary (HMXB) LMC\,X-4 and the foreground star AB\,Doradus.
Besides the known EXOSAT transient EXO053109-6609 which was detected again by ROSAT in 
outburst (\cite{fhaberl-E3:hdp95}), two new HMXBs (the Be/X-ray pulsar RX\,J0529.8-6556, 
\cite{fhaberl-E3:hdpr97} and the OB supergiant system RX\,J0532.5-6551, 
\cite{fhaberl-E3:hpd95}) were discovered by ROSAT. Their nature was recently confirmed
by \cite*{fhaberl-E3:ne02}. The EPIC instruments 
cover all three HMXBs simultaneously in their field of view and, therefore, this region
was selected for deep observations in the guaranteed time programme of XMM-Newton.

\section{The XMM-Newton observation}

The LMC deep field observation was performed by XMM during satellite revolution 152.
The total of 60 ks observing time was split into two parts (ids 0104060101 and 
0104060201) interrupted by a gap of about 90 minutes (start and end of the exposures 
vary somewhat between the EPIC instruments. For the analysis the two observations were
merged. The Optical Monitor was not used because of bright star
constraints. The EPIC instruments covered the field around RA (2000) 05h31m20s and Dec
-65\degr57\arcmin38\arcsec\ with $\sim$13\arcmin\ radius.

\begin{figure*}[ht]
  \begin{center}
     \epsfig{file=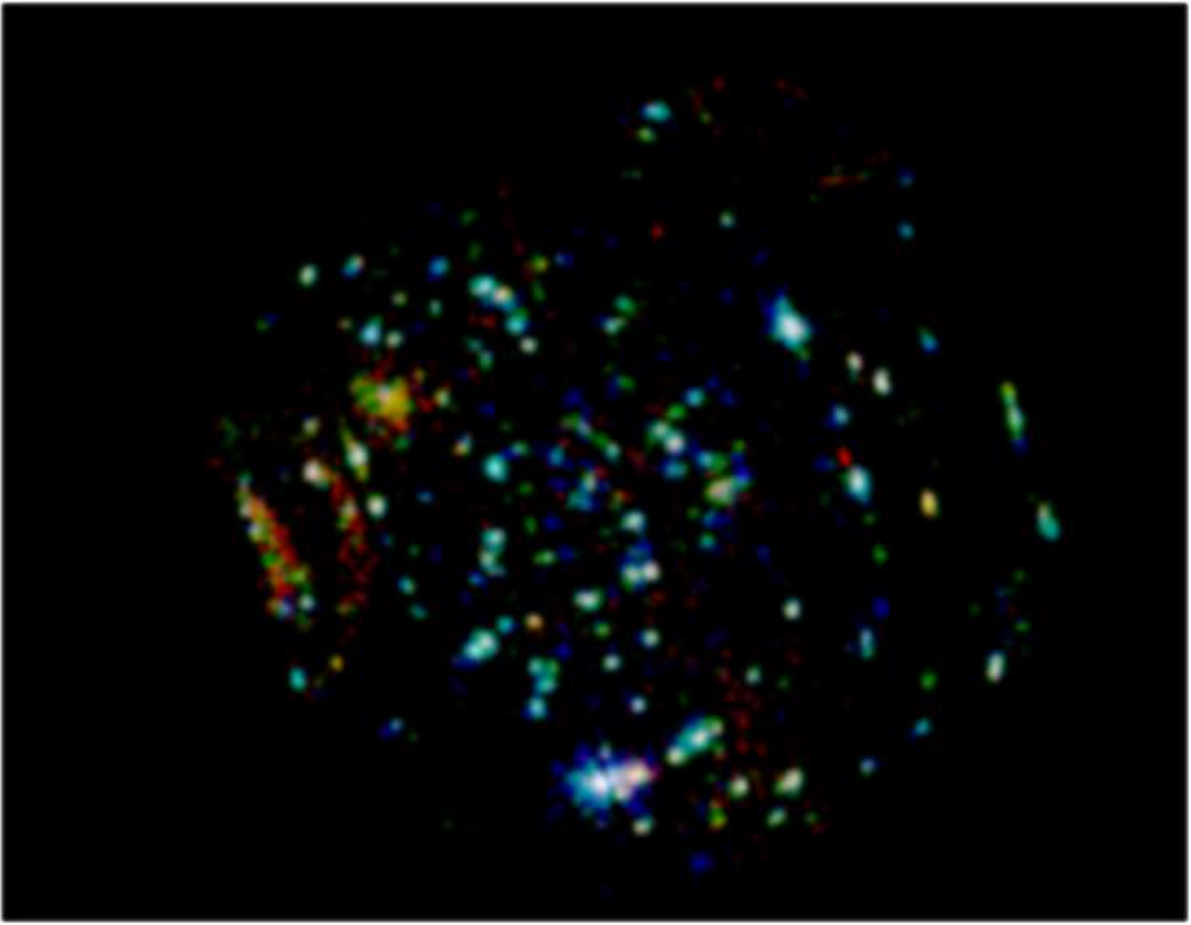, width=16cm}
  \end{center}
\caption{EPIC-pn image of the LMC deep field. The colours represent X-ray intensities in
different energy bands (red: 0.3--1.0 keV, green: 1.0--2.0 keV, blue 2.0--5.0 keV). North
is to the top, east to the left.}  
\label{fhaberl-E3_fig:fig1}
\end{figure*}

Images were created in different energy bands (0.3--1.0 keV, 1.0--2.0 keV, 2.0--5.0 keV)
and combined to a colour image as shown in Fig.~\ref{fhaberl-E3_fig:fig1} for the 
EPIC-pn camera. The broad arc-shaped structures in the east of the image are caused by
X-rays single-reflected by the mirrors from the nearby bright supernova remnant (SNR) N63A. 
A preliminary source detection analysis of the broad energy band revealed about 
150 X-ray sources down to fluxes of 10$^{-14}$ erg cm$^{-2}$ s$^{-1}$. A more detailed 
analysis is, however, still required to produce a reliable source catalogue from this 
field. ROSAT detected about 35 sources in the field.

\subsection{Nature of X-ray sources}

An extended source just NW of the arcs, marked by similar colours, is immediately 
conspicuous as new
SNR candidate. Another, weaker SNR candidate is located NE of the image centre. Both of
the new SNR candidates were also detected by ROSAT and are found in the PSPC catalogue
of the LMC as extended sources (PSPC sources 190 and 205). The lower statistical quality
of the PSPC data however, resulted in a low extent likelihood and did not allow the 
classification as SNR candidates by \cite*{fhaberl-E3:hp99b}.

\begin{figure*}[ht]
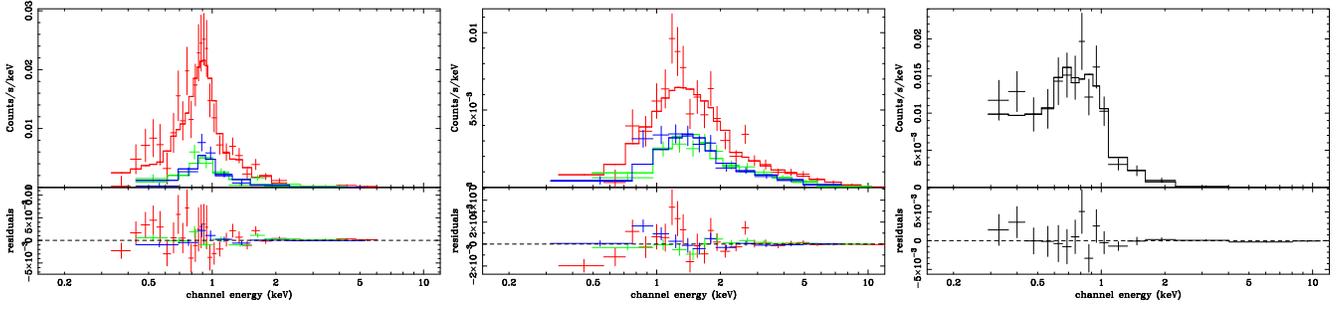

  \begin{center}
     \epsfig{file=fhaberl-E3_fig2a.ps, angle=-90, clip=, width=58mm}
     \epsfig{file=fhaberl-E3_fig2b.ps, angle=-90, clip=, width=58mm}
     \epsfig{file=fhaberl-E3_fig2c.ps, angle=-90, clip=, width=58mm}
  \end{center}
\caption{
EPIC spectra of different types of X-ray sources in the deep LMC field 
(red, black: pn, green: MOS1, blue: MOS2). The best fit models are shown as histograms.
Left: The spectra of a new candidate SNR are well represented by a 
thermal plasma emission model (VMEKAL) with a temperature of 0.71 keV.
Middle: EPIC spectra of a new candidate AGN, best fit by a power-law with photon index 
1.86 and absorbing column density of 5.5{$\cdot 10^{21}$} cm{$^{-2}$}.
Right: EPIC-pn spectrum of a foreground star. 
The spectrum can be represented by a VMEKAL model with a temperature of 0.63 keV and 
little absorption.}  
\label{fhaberl-E3_fig:fig2}
\end{figure*}

The majority of sources appear as blue in Fig.~\ref{fhaberl-E3_fig:fig1} which indicates
a hard X-ray spectrum. These may either be hard X-ray binaries or AGN, the latter 
considerably absorbed by the interstellar gas of the LMC. In the restricted 0.1--2.4 
keV ROSAT energy band X-ray binaries and AGN could not be distinguished from their 
spectrum using hardness ratio criteria (\cite{fhaberl-E3:hp99b}). At least for the 
brighter sources the high statistical quality of the XMM-Newton data allows to 
create broad band, 0.1-- 12 keV spectra. A typical example for an AGN
candidate is a source W to the arcs (also detected by ROSAT PSPC, 202). 
It exhibits a strongly absorbed power-law spectrum with a photon index typical for an 
AGN. Foreground stars appear yellow in the image, their X-ray spectra show low
absorption and little emission above 2 keV.
Example X-ray spectra for one of the new candidate SNRs, AGN and a foreground star 
are plotted in 
Fig.~\ref{fhaberl-E3_fig:fig2}. The 0.1--12 keV energy coverage of the EPIC instruments
easily allows to differentiate between these three types of X-ray sources and to identify
their nature. In particular power-law spectra can be easily distinguished from spectra of
thermal plasma emission with temperatures up to a few keV.

Additional information about the source nature can be obtained from correlations with 
other wavelength bands in order to identify possible counterparts. Already for 
the currently achieved accuracy in the X-ray source positions (systematic 
uncertainties are $\sim$3\arcsec) finding charts produced from Digitized Sky Survey 
images allow to identify optical counterparts for foreground stars and HMXBs in the LMC 
which are both relatively bright in the optical. 
Fig.~\ref{fhaberl-E3_fig:fig3} shows the finding charts for the AGN and foreground star 
as identified from their X-ray spectra in Fig.~\ref{fhaberl-E3_fig:fig2}. While a bright 
stellar object is found in the error circles of the foreground star, no obvious counterpart
is visible for the AGN, consistent with their proposed nature.

\begin{figure}[ht]
  \begin{center}
     \epsfig{file=fhaberl-E3_fig3a.ps, clip=, width=40mm}
     \epsfig{file=fhaberl-E3_fig3b.ps, clip=, width=40mm}
  \end{center}
\caption{DSS2 (R) images with X-ray error circles obtained from the three EPIC 
instruments (green: pn, blue: MOS). The red circle marks the ROSAT PSPC position.
On the left the X-ray position of the AGN candidate is shown. No bright counterpart is
visible in the error circles. On the right a bright object is found in the error circles.
Together with the X-ray spectrum the identification with a foreground star is obvious.}  
\label{fhaberl-E3_fig:fig3}
\end{figure}

\begin{figure}[ht]
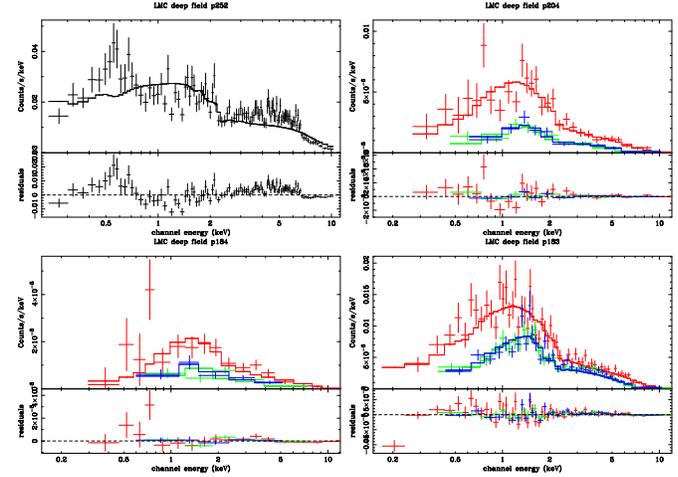

  \begin{center}
     \epsfig{file=fhaberl-E3_fig4a.ps, angle=-90, clip=, width=43mm}
     \epsfig{file=fhaberl-E3_fig4b.ps, angle=-90, clip=, width=43mm}
     \epsfig{file=fhaberl-E3_fig4c.ps, angle=-90, clip=, width=43mm}
     \epsfig{file=fhaberl-E3_fig4d.ps, angle=-90, clip=, width=43mm}
  \end{center}
\caption{EPIC spectra of the three known HMXBs in the field. The Be/X-ray pulsars 
EXO053109-6609 (left) and RX\,J0529.8-6556 (right) are shown on the top. The OB supergiant
system RX\,J0532.5-6551 (lower left) showed the lowest luminosity of {$1.1\cdot10^{34}$} 
erg s{$^{-1}$}. 
The spectrum of EXO053109-6609 is more complicated than a simple power-law model 
and the presence of a soft component is indicated. The spectra of RX\,J0529.8-6556 and 
RX\,J0532.5-6551 are well represented by absorbed 
power-law models with photon index $\sim 1.3$, flatter than for a typical AGN.
A new candidate HMXB (ROSAT PSPC source 183) is suggested by the similar spectrum with 
photon index 1.1 (bottom right).}  
\label{fhaberl-E3_fig:fig4}
\end{figure}

Among the X-ray brightest sources in the investigated field are HMXBs. They usually show
power-law spectra with photon indices typically between 1.1 and 1.5, 
harder than most AGN (Fig.~\ref{fhaberl-E3_fig:fig4}). Moreover the high mass, 
early type star in these accretion powered binary systems is sufficiently luminous to 
be seen as bright star with typical magnitudes of 12--15. 
Figure~\ref{fhaberl-E3_fig:fig5} presents the DSS images for the three known HMXBs with
the error circles obtained from the EPIC instruments over-plotted. Also shown are the 
ROSAT PSPC error circles. The hard power-law spectrum of ROSAT source 183 and the presence
of a possible counterpart with B = 14.2 mag strongly suggest this source to be a HMXB.

\begin{figure}[ht]
  \begin{center}
     \epsfig{file=fhaberl-E3_fig5a.ps, clip=, width=40mm}
     \epsfig{file=fhaberl-E3_fig5b.ps, clip=, width=40mm}
     \epsfig{file=fhaberl-E3_fig5c.ps, clip=, width=40mm}
     \epsfig{file=fhaberl-E3_fig5d.ps, clip=, width=40mm}
  \end{center}
\caption{DSS2 (R) images with X-ray error circles obtained from the three EPIC 
instruments (green: pn, blue: MOS). The red circle marks the ROSAT PSPC position.
EXO053109-6609 (top left) is located on a CCD gap in the pn detector which causes 
the source split into two parts with distorted positions. The nature of the Be/X-ray pulsar
RX\,J0529.8-6556 (top right) and the OB supergiant system RX\,J0532.5-6551 (bottom left) 
were recently confirmed by 
Negueruela \& Coe (2002) using optical spectroscopy. 
Based on the X-ray spectrum and the possible optical counterpart found on the DSS image,
which is of similar brightness than the counterparts of the known HMXBs in the LMC, a new
candidate HMXB is found (bottom right).}  
\label{fhaberl-E3_fig:fig5}
\end{figure}

\begin{figure}[ht]
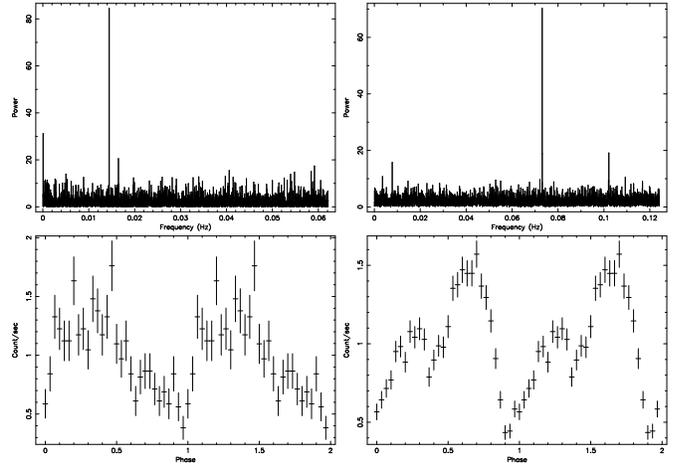

  \begin{center}
     \epsfig{file=fhaberl-E3_fig6a.ps, angle=-90, clip=, width=43mm}
     \epsfig{file=fhaberl-E3_fig6b.ps, angle=-90, clip=, width=43mm}
     \epsfig{file=fhaberl-E3_fig6c.ps, angle=-90, clip=, width=43mm}
     \epsfig{file=fhaberl-E3_fig6d.ps, angle=-90, clip=, width=43mm}
  \end{center}
\caption{Fourier power spectra and folded light curves of the Be/X-ray pulsars 
RX\,J0529.8-6556 (left) and EXO053109-6609 (right). The almost continuous observations
with high sensitivity allow to detect pulsations from HMXBs in the Magellanic Clouds 
at flux levels (0.2--10.0 keV) of {$\sim8\cdot10^{-14}$} erg cm{$^{-2}$} s{$^{-1}$}.}  
\label{fhaberl-E3_fig:fig6}
\end{figure}

A timing analysis was performed for the known HMXBs and the new candidate. For the two
Be/X-ray pulsars pulse periods of 13.667 s (EXO053109-6609) and 69.23 s (RX\,J0529.8-6556) 
were derived, the latter confirming the ROSAT discovery as X-ray pulsar
(Fig.~\ref{fhaberl-E3_fig:fig6}). In this first preliminary
analysis no significant pulsations were found for RX\,J0532.5-6551 and for the new 
HMXB candidate.

\begin{figure}[ht]
  \begin{center}
     \epsfig{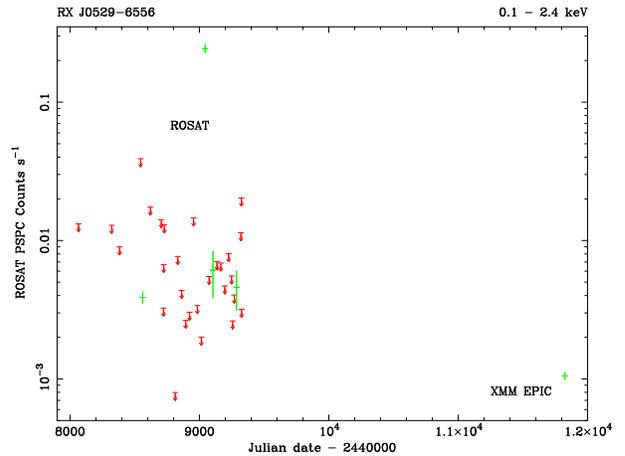}
  \end{center}
\caption{Long term light curve of RX\,J0529.8-6556. For the EPIC point a conversion factor
of 1.0 PSPC counts s{$^{-1}$} per {$1.5\cdot10^{-11}$} erg cm{$^{-2}$} s{$^{-1}$}
was used (assuming the spectral model used to fit the EPIC spectra).}  
\label{fhaberl-E3_fig:fig7}
\end{figure}

The long term light curve of RX\,J0529.8-6556 during the ROSAT observations between
1990 and 1994 was presented by \cite*{fhaberl-E3:hdpr97}. Figure~\ref{fhaberl-E3_fig:fig7}
shows this light curve with the XMM-EPIC point added. The flux during the XMM-Newton
observation was the lowest ever detected from this source and a factor of $\sim$250 lower
than during the outburst in one of the ROSAT observations when the X-ray
pulsations were discovered. 

\subsection{X-ray source classification}

\begin{figure}[ht]
  \begin{center}
     \epsfig{file=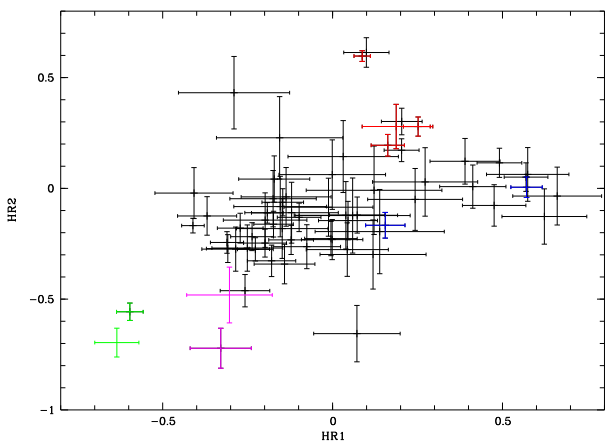, angle=-90, clip=, width=80mm}
  \end{center}
\caption{Hardness ratios of a subsample (errors on HR1 and HR2 less than 0.2) of sources
detected in the field. Sources of known type are marked with different colours: foreground
stars (green), SNRs (pink), AGN (blue) and HMXBs (red). }  
\label{fhaberl-E3_fig:fig8}
\end{figure}

The sample of brighter sources observed in the deep LMC field can be used to define
distinct source properties which allow to classify sources down to lower flux levels. The
classification of the sources according to their X-ray properties provides useful
information for subsequent identification work. As demonstrated with the colour image of
Fig.~\ref{fhaberl-E3_fig:fig1} and the X-ray spectra of the brighter sources, ``X-ray
colours" derived from ratios of count rates in different energy bands (hardness ratios)
can be used to distinguish sources with different emission mechanisms and absorption
column densities. Hardness ratios were defined as HR1 = (B+A)/(B-A) and 
HR2 = (C+B)/(C-B) with A, B and C as count rates in the energy bands
0.3--1.0 keV, 1.0--2.0 keV and 2.0--5.0 keV, respectively. Given the
foreground absorption to the observed part of the LMC and intrinsic LMC and source 
absorption the observed column densities derived from the spectra are between
$\sim$6{$\cdot 10^{20}$} cm{$^{-2}$} and $\sim$6{$\cdot 10^{21}$} cm{$^{-2}$}. 
Such column densities mainly affect the spectrum below 1 keV and therefore 
HR1, while HR2 characterizes the source intrinsic spectrum above 1 keV. In
Fig.~\ref{fhaberl-E3_fig:fig8} HR1 is plotted versus HR2 for those sources with error on
both hardness ratios of less than 0.2. HR1 is directly correlated with the absorption  
column densities and foreground stars are therefore found on the left side of the diagram
while AGN, absorbed by the galactic and the LMC interstellar absorption are expected to be located
towards the right side. Most of the unknown sources found with HR1 values larger than
$\sim0.2$ are most likely AGN. Their hardness ratios are compatible with absorbed power-law 
spectra with photon index of $\sim$1.8. The HMXBs exhibit the hardest X-ray spectra and 
highest values of HR2. A large number of unknown sources is found with HR1 below 0 which 
implies absorption column densities below the sum of galactic and LMC absorption. To be 
explained as AGN a soft spectral component would be required. However from log N -- log S
considerations a smaller AGN contribution is expected. The sensitivity limit 
of the observation (the faintest sources detected have fluxes 
$\sim$3.7$\cdot 10^{-15}$ erg cm{$^{-2}$} s{$^{-1}$} which correspond to luminosities of 
$\sim$1.1$\cdot 10^{33}$ erg s{$^{-1}$} at the distance of the LMC, 50 kpc) should allow to detect
the brightest cataclysmic variables (CVs) in the LMC. However it remains unclear if CVs 
and low mass X-ray binaries (LMXBs) can explain the large number of unknown sources.

\section{Summary}

In a preliminary analysis of a 60 ks XMM-Newton observation, pointed to an area
in the Large Magellanic Cloud, new candidates for supernova remnants and a high mass
X-ray binary were discovered. The high sensitivity and broad energy band of the
EPIC instruments allows to determine the type of the X-ray sources to much lower flux levels
as was possible before. While most of the high mass X-ray binaries in the Magellanic Clouds
so far were discovered during X-ray outburst these systems can now be detected and 
identified during periods of quiescence. This will allow to reach a more complete census of
this source population in our neighboring galaxies.

First results revealed the presence of a relatively large number of unknown sources, too soft
to be explained by AGN with power-law spectra absorbed through the LMC. Further 
investigations are needed to see if they can be explained by either background 
AGN with soft excess or sources intrinsic to the LMC like low
mass X-ray binaries or cataclysmic variables.

\begin{acknowledgements}

The XMM-Newton project is an ESA Science Mission with instruments and
contributions directly funded by ESA Member States and the USA (NASA). The
XMM-Newton project is supported by the Bundesministerium f\"ur Bildung und
For\-schung / Deutsches Zentrum f\"ur Luft- und Raumfahrt (BMBF / DLR), the
Max-Planck-Gesellschaft and the Heidenhain-Stif\-tung.

The finding charts are
based on photographic data obtained using The UK Schmidt Telescope.
The UK Schmidt Telescope was operated by the Royal Observatory
Edinburgh, with funding from the UK Science and Engineering Research
Council, until 1988 June, and thereafter by the Anglo-Australian
Observatory.  Original plate material is copyright (c) the Royal
Observatory Edinburgh and the Anglo-Australian Observatory.
The plates were processed into the present compressed digital form with
their permission.  The Digitized Sky Survey was produced at the Space
Telescope Science Institute under US Government grant NAG W-2166.

\end{acknowledgements}

\end{document}